# Chronopixel Vertex Detectors for Future Linear Colliders

*C. Baltay, W. Emmet, and D. Rabinowitz*
*Yale University, New Haven, CT 06520, USA*
*J. Brau, N. Sinev, and D. Strom*
*University of Oregon, Eugene, OR 97403*

## Abstract

Over the past few years we have developed, in collaboration with the SARNOFF Corporation, a design of a Monolithic CMOS Pixel Detector for ILC Vertex Detectors. The unique feature of this design is that each hit is accompanied by a time tag with sufficient precision to assign each hit to a particular bunch crossing of the ILC (thus the name Chronopixel). This reduces the occupancy even in the innermost Vertex Detector layer to negligible levels, allowing a robust Vertex Detector even in the case that the backgrounds in the ILC are higher than presently estimated.

## 1. Introduction

Studies carried out in the U.S., Europe, and Asia, have demonstrated the power of a pixel vertex detector in physics investigations at a future high energy linear collider. At one time, silicon CCD's (Charged Coupled Devices)[1] seemed like the detector elements of choice for vertex detectors for future Linear $e^+ e^-$ Colliders. However, with the decision for a cold TESLA-like superconducting technology for the future International Linear Collider (ILC), the usefulness of CCD's for vertex detection has become problematical. The time structure of this cold technology is such that it necessitates an extremely fast readout of the vertex detector elements and thus CCD's as we know them will not be useful. New CCD architectures are under development[2] but have yet to achieve the required performance. For these reasons there is an increased importance on the development of Monolithic CMOS pixel detectors that allow extremely fast non sequential readout of only those pixels that have hits in them. This feature significantly decreases the readout time required. Recognizing the potential of a Monolithic CMOS detector, we initiated an R&D effort to develop such devices[3]. Another important feature of our present conceptual design for these CMOS detectors is the possibility of putting a time stamp on each hit with sufficient precision to assign each hit to a particular bunch crossing. This significantly reduces the effective backgrounds in that in the reconstruction of any particular event of interest we only need to consider those hits in the vertex detectors that come from the same bunch crossing.

The detailed design of these devices has been completed by SARNOFF and checked by SPICE Simulations. The fabrication of the first set of prototype devices has been completed and packaged. We have developed a detailed testing plan and have built the test electronics at SLAC. The testing of the first prototypes has been completed, showing that the general concept of the devices is working, but there were a number of design errors that need correction. The design of the second set of prototypes is now underway and we expect to be testing these in the next six months.

## 2. General Description of the Chronopixel Design

In our design we are assuming an ILC beam structure with 2820 bunches in a bunch train with 330 nanosec between bunches and 200 millisec between bunch trains.

The current Chronopixel design is for chips up to 12.5 cm x 2.0 cm in size with a single layer of 10 μ m x 10 μ m charge sensitive pixels. Each pixel has its own electronics under it, but both the sensitive layer and the electronics are made of one piece of silicon (monolithic CMOS) which can be thinned to a total thickness of 50 to 100 μ m, with no need for indium bump bonds. The electronics for each pixel will detect hits above an adjustable threshold. For each hit the time of the hit is stored in each pixel, up to a total of four different hit times per pixel, with sufficient precision to assign each hit to a particular beam crossing (thus the name "chronopixels" for this device). Hits will be accumulated for the 2820 beam crossing of a bunch train and the chip is read out during the 200 millisec gap between bunch trains. There is sufficient intelligence in each pixel so that only pixels with one or more hits are read out, with the x, y coordinates and the time t for each hit. With 10 micron size pixels we do not need analog information to reach a 3 to 4 micron precision so at the present we plan on digital read out, considerably simplifying the read out electronics.

To get some feeling for the hit rates and occupancies we use the estimated 0.03 hits/mm$^2$/beam crossing for the worst case innermost layer[4]. With 2500 mm$^2$ per chip (a total of 25 x 10$^6$ pixels/chip) and 2820 beam crossings per train we expect 2 x 10$^5$ hits/chip/bunch train, or an occupancy of the order of one percent.

This appears much too high to allow efficient pattern recognition. The crucial element of our design is the availability of the time information (i.e., bunch crossing number) with each hit. If we trigger on an event that we are interested in from another part of the detector (tracker or calorimeter) with a time, i.e., the bunch crossing number known, we need to look only at those vertex detector hits which are consistent in time with the event of interest and the beam induced occupancy drops to below $10^{-5}$ per pixel (SLD worked well with an occupancy of ~ $10^{-3}$ per pixel in the Vertex Detector).

This design has been described in various conference proceedings (J. Brau at Bangalore LCWS 2006, C. Baltay at SLAC-Novosibirsk Instrumentation Conference, April, 2006, and J. Brau at the Hiroshima Semiconductor Detector Conference, Carmel, September, 2006, ALCPG08, Fermilab, N.Sinev LCWS08, Chicago). A detailed description of this design is documented in two reports by SARNOFF, "HEP Vertex Detector Macropixel Design," February 28, 2006, and HEP Vertex Detector Chronopixel Array Design," January 31, 2007.

## 3. Detailed Design

SARNOFF has carried out a design of the electronics under each pixel of this chronopixel array. A block diagram of the electronics in each pixel is shown in Figure 1. The functionality of this design has been verified by an hspice simulation.

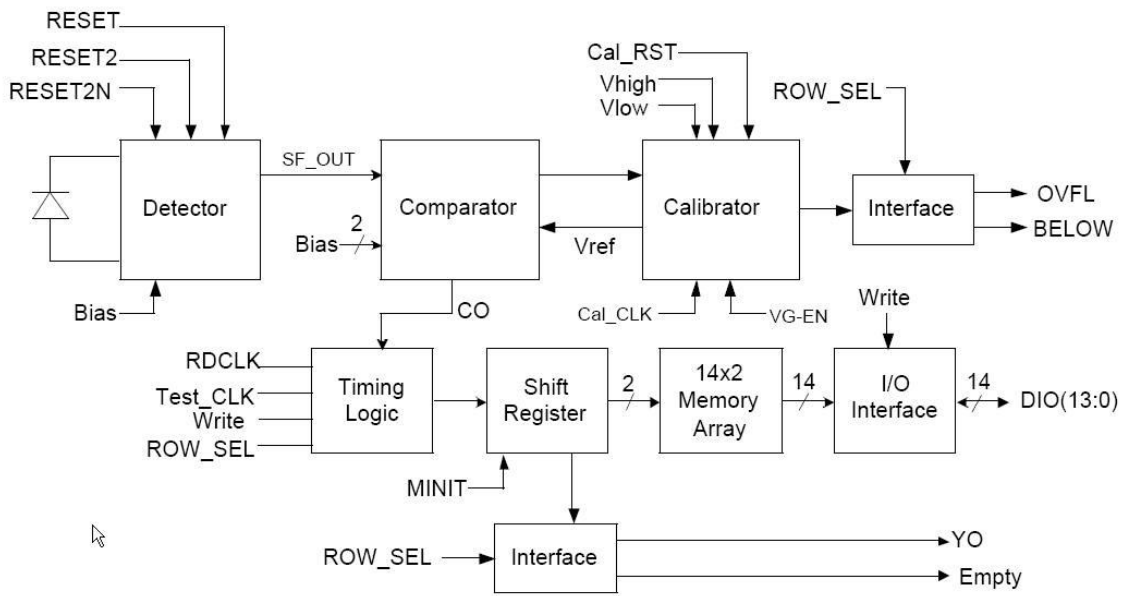

Figure 1: Block Diagram of a single pixel

### 3.1. Chronopixel Array Design

The architecture of each pixel of the chronopixel array is shown in Figure 1. After each bunch crossing, the signal in each pixel is compared to a preset, calibrated threshold level. If the signal is above threshold, the time of the bunch crossing is stored in the first slot of the 14 bit, 4 deep timing array. A hit exceeding threshold from a subsequent bunch crossing is stored in the second time slot, and so on up to four hit times in a bunch train. The poisson probability of exceeding 4 hits per hit pixel pixel in a train of 2820 bunches is less than $10^{-4}$. The time of the hits for the pixels with hits in them are read out in the 200 msec gap between bunch trains. The detector sensitivity is 10 μ V per electron and the detector noise is estimated to be 25 electrons. The comparator accuracy is 0.2 mV rms. To accomplish all this each pixel has 645 transistors in it, as shown in Figure 2. To squeeze all of this into a 10 μ x 10 μ pixel will require the 45 nanometer process technology which is expected to be available in the next few years.

To reduce the error on the comparator accuracy to below 0.2mV each pixel will have a self-calibrating threshold. In the prototype we will try two options for the variable threshold circuit. Half of the pixels will be instrumented with a circuit that has 8 values of the threshold. The second half of the pixels will have a fine and course threshold for a total 32 different threshold values.

Before each bunch train arrives at the detector, the threshold will be scanned in each pixel to locate the zero of the comparator. Shift registers will be used to store the effective zero of the comparator. A common offset will then be applied to all channels to put the threshold approximately 5 sigma above zero.

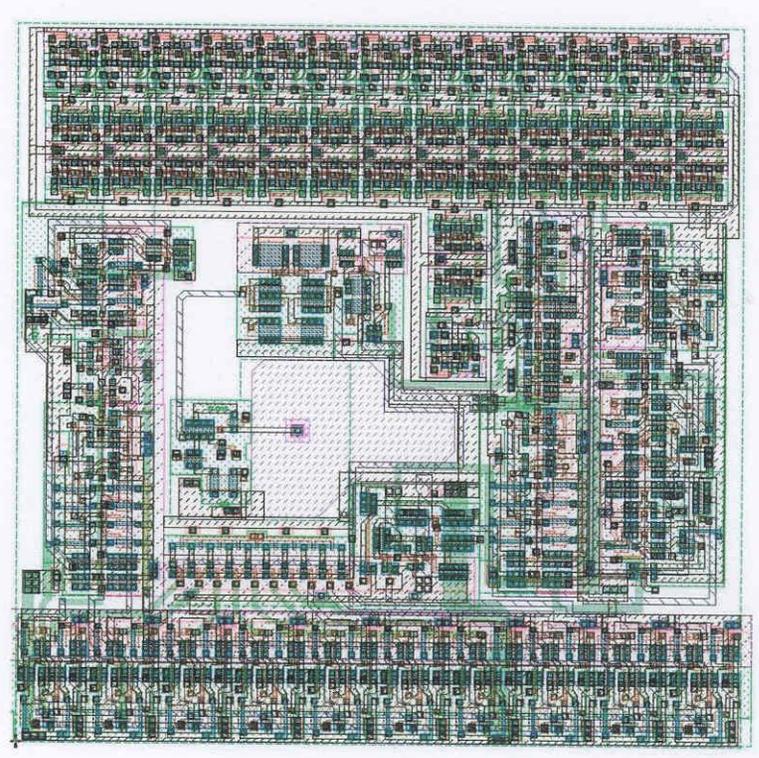

Figure 2: Layout of a single Pixel

### 3.2. Signal to Noise Issues

A simulation to estimate the energy loss, in units of electrons, of 1 Gev pions in various thicknesses of silicon has been carried out by Su Dong from SLAC. This energy loss has a broad distribution. The peak or mean energy loss, however, is not the relevant consideration. The question to ask is above what signal level do we detect 99% of the particle crossings. These are shown in column 2 of Table 1 for various silicon epilayer thicknesses. We then want to set the threshold for the signal in each pixel so that we get 99% efficiency even if the charge is shared by two adjacent pixels. These numbers are shown in column 3 of Table 1. To keep the number of fake hits (due to noise fluctuating above the threshold) below the .03 background hits/pixel/beam crossing the threshold should be 5 sigma above the noise, i.e. the maximum acceptable noise per pixel has to be below one fifth of the threshold. These noise levels are given in the last column of Table 1.

Our design is based on a 15 micron thick epitaxial layer, where we expect to set the threshold at 200 electrons/pixel. The estimated noise of 25 electrons in our design is comfortably below the acceptable 40 electron noise level.

**Table 1: Signal to Noise Considerations**

| Epilayer Thickness (microns) | Electrons at 99% Efficiency | Threshold (electrons) | Acceptable Noise (electrons) |
|---|---|---|---|
| 4 | 40 | 20 | 4 |
| 7 | 125 | 63 | 13 |
| 10 | 250 | 125 | 25 |
| 15 | 400 | 200 | 40 |
| 20 | 550 | 275 | 55 |

### 3.3. Read Out Scheme

Each chip will consist of 2000 columns with 12500 pixels each. Each chip will be divided into 40 read out regions of 50 columns each. At the end of the bunch train when the electromagnetic interference due to the beam has died off the 40 read out regions will be read out in parallel at 25 MHz into a FIFO buffer located at the end of each chip. The contents of the FIFO buffer will be read out off the chip at 1 GHz. We thus expect to read out the full chip ($2 \times 10^5$ hits, with 38 bits per hit) in about 8 millisec. This leaves a safety margin of 25 with the 200 millisec gap between trains.

### 3.4. Power Consumption

The power consumption of this circuitry has been estimated. The analog parts of the circuit consume most of the power, estimated at this stage of the design to be ~ 15 milliwatts/mm$^2$. The remaining digital components are estimated to be around 0.05 milliwatts/mm$^2$. The analog components are only needed during the time when hits are accumulated during the bunch train, ~ 1 millisec. The average power can thus be reduced by a factor of ~ 100 by turning off the analog parts during the 200 millisec digital readout. This would reduce the average power consumption to the vicinity of 0.4 watts per chip or to the order of 100 watts for the vertex detector, which seems acceptable.

### 3.5. Charge Spreading

In order to use digital readout the charge spreading has to be kept well below the pixel size. This is aided in this design by depleting the epitaxial layer to the maximum extent possible. Maximal depletion of the 15 micron thick epilayer at the voltages allowed in the 45 nm process will benefit from the highest feasible silicon resistivity. Such depletion combined with diffusion should adequately contain the charge spreading. Another factor relevant in this respect is the large Lorentz angle if the Vertex Detector operates in a large axial magnetic field.

### 4. Design and Fabrication of the First Set of Prototypes

As mentioned above, the ultimate Chronopixel design with 10 micron x 10 micron pixels requires a 45 nanometer process technology. We expect that this technology will be available in the next few years, in plenty of time for fabrication of these detectors. However, for the fabrication of the first Chronopixel prototypes the 45 nm technology was not easily available (and very expensive). We therefore decided to fabricate the first set of prototypes with 50 micron x 50 micron pixels with 180 nm technology by the TSMC foundry in Taiwan. The assumption is that the pixel size scales with the feature size of the technology. The fact that all of this electronics could be fit into a 50x50 micron pixel in the 180 nm technology implies then that with the 45 nm technology we can reduce the pixel size to 12.5 microns. We hope that with improvements in the design we will be able to achieve 10 micron pixel size, although 12.5 micron pixels would also be quite acceptable.

Another important issue was the resistivity of the silicon. In order to fully deplete the 15 micron thick epitaxial layer to keep the charge spreading small (see section 3.5 above) we will need high resisitivity silicon, in the vicinity of 10 kohm cm, in the final devices. However at the time that we started prototype 1 such silicon was available on special order only and would have considerably increased both the cost and the time required to produce the first prototypes. We therefore decided to use silicon readily available at TSMC lot runs with a 7 micron thick epilayer with 10 ohm cm resistivity. Thus the epilayer in the first prototype will not be fully depleted.

We felt that these were sensible compromises, making the first prototype affordable with the available funding on a reasonably short time scale. The main purpose of this prototype was to test the electronics performance of the Chronopixel design i.e., noise performance, comparator accuracy and stability, scan speed and power dissipation. We are carrying out 3 dimensional simulations to estimate other features of the prototype such as charge collection efficiency and charge spreading with the given pixel and epilayer parameters and will compare the actual performance of the prototypes with these estimates. This will provide valuable information for the design of the second prototype.

The design and fabrication of the first set of prototypes have been completed by SARNOFF and the TSMC fab house. Figure 3 shows a photograph of these devices.

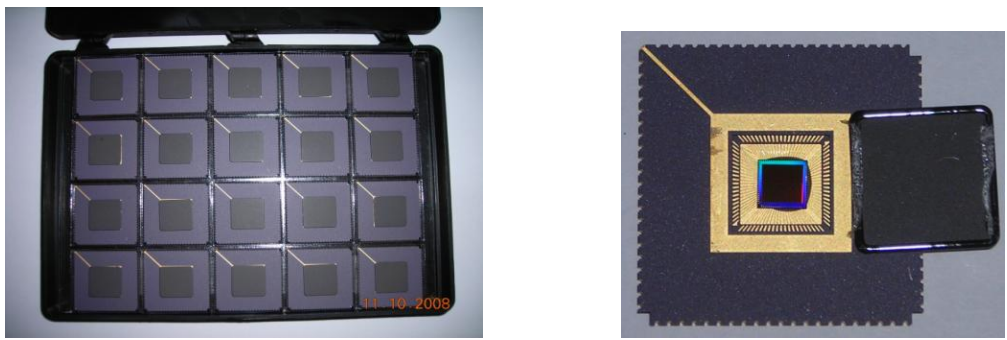

Figure 3. Photographs of the packaged Prototype 1 devices.

## 5. Results of the Prototype 1 Tests

The testing of prototype 1 have been completed. The tests showed that the general concept of the device is working. There were some errors in the power distribution on the chips which led to only a portion of the chip to be operational. These errors have been diagnosed and will be corrected in the second prototype. The calibration circuit works as expected in the test pixels. The noise figures with the soft reset are within specification (measured noise of 24 electrons, where the specs were 25 electrons). The sensitivity was measured to be 36 microvolts per electron, better then the design sensitivity of 10 moicrovolts per electron. The comparator accuracy was three times worse then the specification. This needs to be improved in prototype 2. The sensor leakage current was measured to be $1.8 \times 10^{-8}$ Amps/cm$^2$ which is fine. And finally the readout time was satisfactory.

## 6. Plans for Prototype 2

Encouraged by the results of the first prototype, we proceeded to the design of the second set of prototypes. We are planning on using the TSMC 90 nanometer technology, which will allow us to reduce the pixel size to 25 x 25 microns. For reasons of both cost and speedier fab time we will use standard silicon wafers with a 7 micron epilayer and low resistivity. We obviously plan to correct the shortcomings of the first prototypes in this prototype. In addition we plan to go to the NMOS technology which promises an improved charge collection efficiency.

The design of the second prototype is now quite far along and we expect the fabrication of these prototypes to start toward the end of this calendar year, and we plan on testing them early next year.